\documentclass[twocolumn,showpacs,preprintnumbers,amsmath,amssymb,showkeys]{revtex4}

\usepackage{graphicx}
\usepackage{dcolumn}
\usepackage{bm}
\usepackage{bezier}

\begin{document}

\title{Persistent currents in quantum phase slip rings}
\author{Andrew G. Semenov$^{1}$
and Andrei D. Zaikin$^{2,1}$
}
\affiliation{$^1$I.E.Tamm Department of Theoretical Physics, P.N.Lebedev
Physics Institute, 119991 Moscow, Russia\\
$^2$Institute of Nanotechnology, Karlsruhe Institute of Technology (KIT), 76021 Karlsruhe, Germany
}

\begin{abstract}
We investigate the effect of interacting quantum phase slips on persistent current and its fluctuations in ultrathin superconducting nanowires and nanorings pierced by the external magnetic flux. We derive the effective action for these systems and map the original problem onto an effective sine-Gordon theory on torus. We evaluate both the flux dependent persistent current and the critical radius of the ring beyond which this current gets exponentially suppressed by quantum fluctuations. We also analyze fluctuations of persistent current caused by quantum phase slips. At low temperatures the supercurrent noise spectrum has the form of coherent peaks which can be tuned by the magnetic flux. Experimental observation of these peaks can directly demonstrate the existence of plasma modes in superconducting nanorings.

\end{abstract}

\pacs{73.23.Ra, 74.25.F-, 74.40.-n}

\maketitle

\section{Introduction}

It is well known that fluctuation effects may play an important role in superconducting nanowires. At low temperatures the most significant
fluctuations are quantum phase slips (QPS) \cite{AGZ,Bezr08,Z10,Bezrbook}. Provided the superconducting wire is sufficiently thin quantum fluctuations may yield temporal local suppression of the absolute value of the superconducting order parameter field $\Delta (x)=|\Delta(x)|e^{i\varphi(x)}$ in different points along the wire. As soon as the modulus of the order parameter $|\Delta(x)|$ in the point $x$  vanishes, the
phase $\varphi(x)$ becomes unrestricted and can jump by the value $\pm 2\pi$. After this process the modulus $|\Delta(x)|$ gets restored, the phase becomes single valued again and the system returns to its initial state accumulating the net phase shift $\pm 2\pi$. According to the Josephson relation $V=\hbar\dot{\varphi}/2e$ each
such QPS event causes either positive or negative voltage pulse. In the presence of an external bias current $I\propto |\Delta |^{2}\nabla\varphi$ "positive" phase slips become more likely than "negative" ones and the net voltage drop $V$ occurs across the wire.
Hence, quantum phase slips may yield nonvanishing resistance $R=V/I$ of superconducting nanowires even at very low temperatures. This effect was observed in a number of experiments with ultrathin superconducting  wires \cite{BT,Lau,Zgi08} (see also \cite{AGZ,Bezr08,Z10,Bezrbook} for a detailed review of relevant experiments).

\begin{figure}[h]
a)\includegraphics[width=0.89\columnwidth]{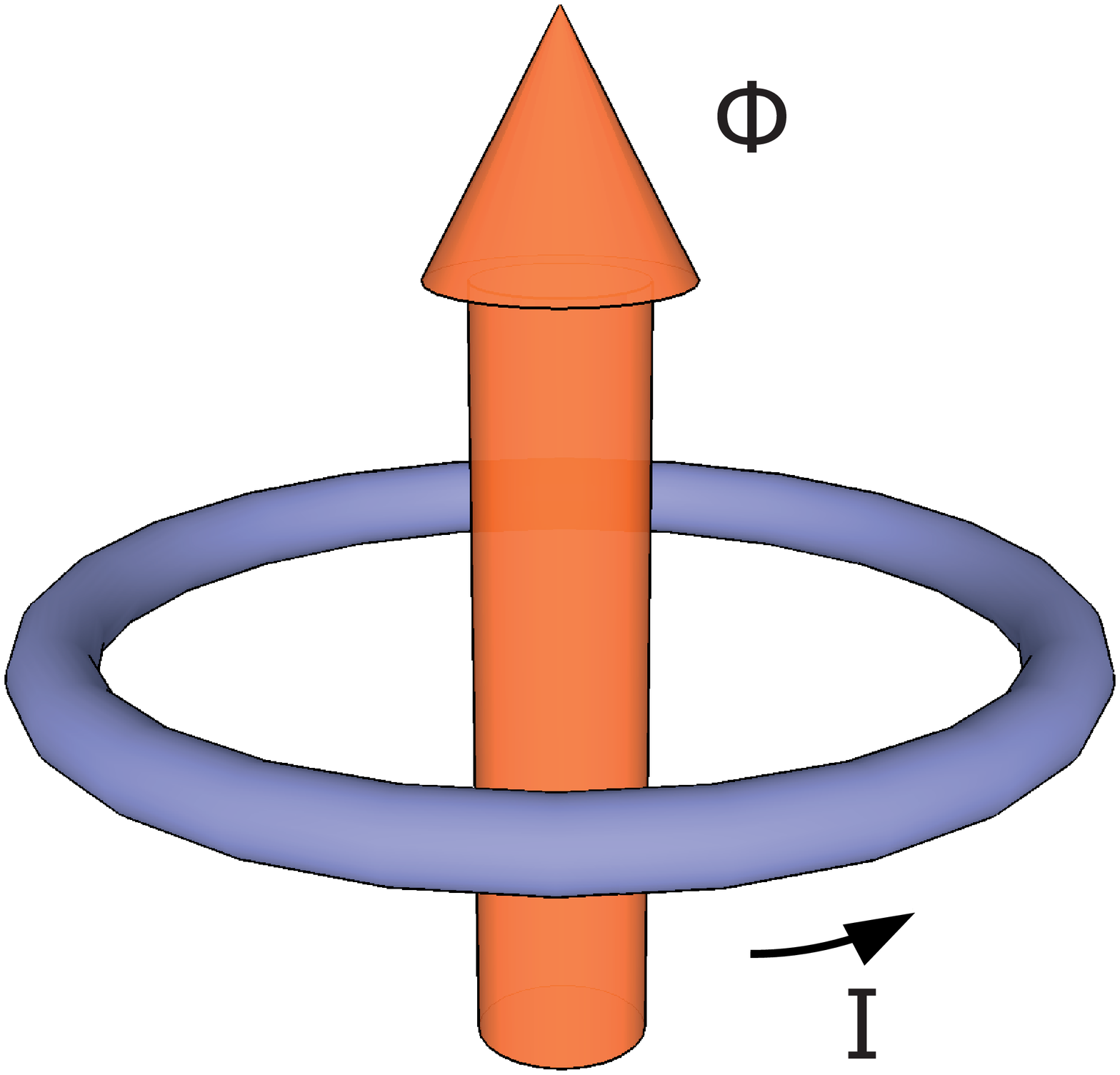}
b)\includegraphics[width=0.89\columnwidth]{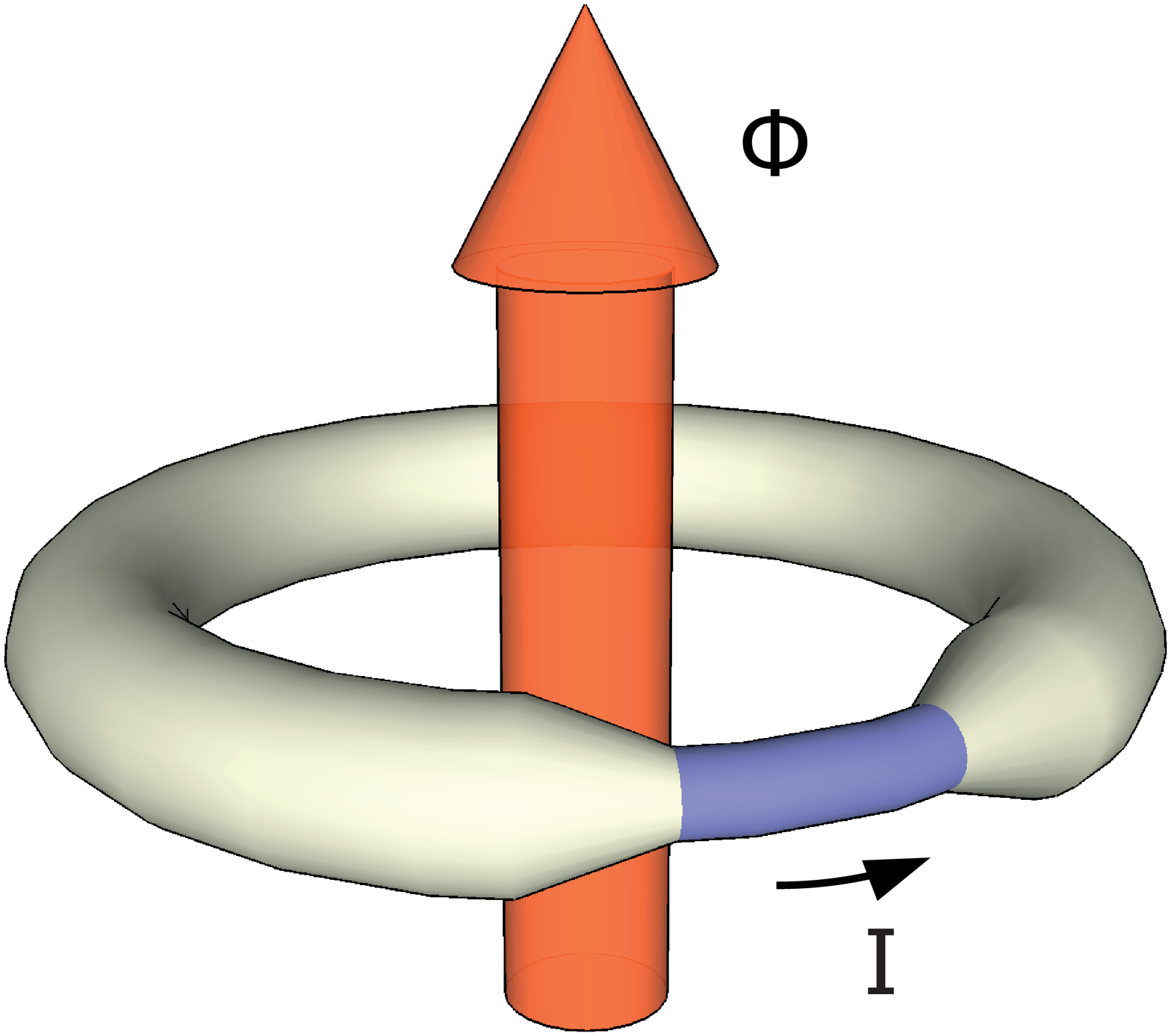}
\caption{(Color online) a) Quantum phase slip ring, i.e. an ultrathin superconducting ring threaded by an external magnetic flux.
b) Quantum phase slip junction embedded in a thick superconducting ring.}
\label{fig1}
\end{figure}

It is also important to emphasize that according to the
existing microscopic theory \cite{ZGOZ,GZQPS} QPS represent
quantum coherent objects which may significantly affect not only transport but also equilibrium ground state properties of superconducting nanowires. One of such fundamental properties of
a quantum coherent ground state is the possibility for persistent currents to flow around superconducting rings threaded by an external
magnetic flux, see Fig. \ref{fig1}a. Provided such a ring becomes sufficiently thin, quantum phase slips proliferate and may drastically modify both
the magnitude of persistent current (PC) and its dependence on the magnetic flux \cite{MLG,AGZ,Z10}. In addition, quantum phase slips
cause non-vanishing PC noise \cite{SZ12} which otherwise would be totally absent in the limit of low temperatures. Coherent nature of quantum phase slips was convincingly demonstrated in recent experiments \cite{AstNature,Ast13}.

All the same phenomena can also occur in a somewhat different
situation of rings consisting of thicker and thinner
parts, as it is shown in Fig. \ref{fig1}b. In this case QPS effects are negligible in a thicker part of the
ring and they can only occur in its thinner part which was named
a {\it quantum phase slip junction} \cite{MN}. The reason for that
is the duality between the properties of these
systems and those of mesoscopic Josephson
junctions: It was argued \cite{MN} that any known result on electron transport in circuits
containing Josephson junctions can be mapped onto a dual
result for a QPS junction in a dual circuit. The same duality property
was earlier predicted for Josephson junctions themselves
\cite{averin,Z90,SZ90} which can also be considered as QPS junctions if one
interchanges the canonically conjugate phase and charge variables describing quantum dynamics of these systems. This duality property can be used in metrology, e.g., for practical implementation of the
electric current standard. It was also proposed \cite{MH} to use rings with QPS junctions for experimental realization of quantum phase slip flux qubits.

Although QPS junctions can indeed -- under certain conditions -- behave similarly to (dual) Josephson junctions, the systems depicted in Fig. \ref{fig1} are clearly not identical to such junctions in several important aspects.
One of them is that a (zero-dimensional) junction is now replaced by a (quasi-one-dimensional) superconducting wire which excitation spectrum is rather different, e.g., due to the presence of plasmon modes \cite{ms}
lacking in small area Josephson junctions. As a result, different quantum phase slip events start "talking" to each other and eventually become correlated due to logarithmic interaction between them both in space and in time \cite{ZGOZ}. This problem of logarithmically interacting gas of phase slips can be routinely mapped onto an effective sine-Gordon model \cite{SZ12,S10,HN} which is exactly dual
to that describing long Josephson barriers.

In this paper we will develop a detailed
analysis of this effect and demonstrate that it can essentially modify ground state properties of generic systems under consideration. As QPS effects are basically the same in uniform superconducting nanorings (Fig. \ref{fig1}a) and QPS
junctions (Fig. \ref{fig1}b), in what follows -- just for the sake of definiteness -- we will merely address the former systems.
In order to distinguish them from thicker superconducting rings where QPS effects are negligible and, on the other hand, to emphasize their similarity to
QPS junctions below we will call such systems {\it quantum phase slip rings}.

The structure of our paper is as follows. In section II we briefly
discuss the effect of PC fluctuations in superconducting rings in the absence of quantum phase slips. In section III we set up our basic formalism deriving both the effective action and the generating functional for PC in superconducting nanorings with quantum phase slips. This formalism is then directly employed in section IV where we
evaluate both the average value of persistent current and PC noise
in quantum phase slip rings with QPS interactions. A brief discussion of our results is presented in Sec. V. Some general expressions and technical details of our derivation are relegated to Appendices A and B.

\section{Superconducting rings without QPS}
Let us start our analysis by considering not very thin superconducting wire in the form of a ring of radius $R$ pierced by the magnetic flux $\Phi_x$. Provided the wire is thick enough one can ignore fluctuations of the absolute value of the order parameter field $|\Delta (x,\tau )|$, where $x$ is the coordinate along the wire and $\tau$ is the imaginary time, $0\geq \tau \geq \beta \equiv 1/T$. In this case QPS effects are strongly suppressed and the partition function of the system $\mathcal  Z$ can be expressed via the following path integral over the superconducting phase variable $\varphi (x,\tau )$ \cite{ZGOZ,GZQPS,ogzb}:
\begin{equation}
\mathcal  Z=\sum\limits_{m,n}\int \mathcal D\varphi e^{-\frac{\lambda}{2\pi}\int dxd\tau\left(v(\partial_x\varphi)^2+v^{-1}(\partial_\tau\varphi)^2\right)},
\label{partf}
\end{equation}
where we defined
\begin{equation}
\lambda =\frac{\pi^{2} N_{0}D\Delta s}{2v},  \quad v=\sqrt{\frac{\pi\sigma \Delta s}{C}}
\end{equation}
as respectively the effective coupling constant \cite{ZGOZ} and
the velocity of low energy plasmon modes (the so-called Mooji-Sch\"on modes \cite{ms}) propagating along the wire.
Here $N_{0}$ is the density of states at Fermi level, $D$ is the diffusion coefficient, $\Delta$ is the mean field value
of the superconducting order parameter in the wire, $s$ is its cross section, $\sigma =2e^2N_0D$ is the Drude normal state
conductivity and $C$ is the wire capacity per unit length.

In order to account for the ring geometry and for the magnetic flux inside the ring it is necessary to
impose proper boundary conditions for the path integral (\ref{partf}). These boundary conditions should include
periodicity of the phase variable $\varphi$ both in imaginary time and in space (after each full rotation around the ring.
On top of that, it is necessary to keep in mind that the phase is defined up to $2\pi m$, where $m$ is an arbitrary integer number
(the so-called winding number). Finally, the boundary conditions should be sensitive to the magnetic flux $\Phi_x$
piercing the ring. Putting all these requirements together, we obtain
\begin{eqnarray}
\varphi(x,0)=\varphi(x,\beta)+2\pi m,\nonumber\quad \\
\varphi(L,\tau)=\varphi(0,\tau)+2\pi(\phi_x+n).
\label{boucond}
\end{eqnarray}
Here $L=2\pi R$ is the ring perimeter, $\phi_{x}=\Phi/\Phi_{0}$ and $\Phi_0=\pi c/e$ is the superconducting
flux quantum. Combining Eqs. (\ref{partf}) and (\ref{boucond}), after a simple calculation one finds
\begin{equation}
\mathcal Z\sim\sum\limits_{n=-\infty}^\infty e^{-\beta E_n(\phi_x)}=\sqrt{\frac{LT}{2\lambda v} }\vartheta_3(\pi\phi_x,e^{-\frac{\pi LT}{2\lambda v}})
\end{equation}
where $E_n(\phi_x)=2\pi\lambda v (n+\phi_x)^2/L$, and $\vartheta_k(u,q)$ is the Jacobi Theta function.

In what follows we will mainly be interested in evaluating the correlation functions for the current flowing around the ring.
This current is expressed via the phase variable as
\begin{equation}
I(\tau)=\frac{2 e v\lambda}{\pi L}\left(\varphi(L,\tau)-\varphi(0,\tau)\right).
\label{current0}
\end{equation}
Employing this relation it is straightforward to demonstrate that all current correlators are time independent and
can be expressed through the derivatives of the Theta function $\vartheta_3^{(k,0)}$ as
\begin{multline}
\langle \hat I(\tau_1)...\hat I(\tau_k)\rangle=
\left(\frac{4ev\lambda}{L}\right)^k\frac{\sum\limits_{n=-\infty}^\infty (n+\phi_x)^ke^{-\beta E_n(\phi_x)}}{\sum\limits_{n=-\infty}^\infty e^{-\beta E_n(\phi_x)}}\\
=\sum\limits_{n=0}^{2n\leq k}\frac{(-eT)^kk!}{n!(k-2n)!}\left(\frac{2\lambda v}{\pi L T}\right)^n\frac{\vartheta_3^{(k-2n,0)}(\pi\phi_x,e^{-\frac{\pi LT}{2\lambda v}})}{\vartheta_3(\pi\phi_x,e^{-\frac{\pi LT }{2\lambda v}})}
\label{curcor}
\end{multline}
In particular, for the expectation value of persistent current $I(\phi_{x})=\langle I(\tau )\rangle$ flowing around the ring Eq. (\ref{curcor}) yields the standard result which in the limit $T \to 0$ further reduces to
\begin{equation}
I(\phi_{x})=\frac{\pi e N_0D\Delta s}{R}\mathrm{min}_{n}\left(  n+\phi_{x}\right).
\label{evcur}
\end{equation}

In order to analyze PC fluctuations let us define the irreducible Matsubara correlator
\begin{equation}
\Pi(\tau)=T\sum\limits_k e^{-i\omega_k\tau}\Pi_{i\omega_k}=\langle \hat I_M(\tau)\hat I_M(0)\rangle-\langle\hat I\rangle^2,
\label{pi}
\end{equation}
where $\hat I_M=e^{\tau\hat H}\hat Ie^{-\tau\hat H}$ is the current operator in the Matsubara representation, $\hat H$ is the system Hamiltonian and $\omega_k=2\pi k T$ is Matsubara frequency. It follows from Eq. (\ref{curcor}) that this correlator does not depend on time and can
easily be evaluated employing the above expressions. It is also important that from the expression for the imaginary time correlator
(\ref{pi}) one can directly recover the real time PC noise power spectrum
\begin{equation}
S_\omega=\int dt e^{i\omega t} S(t),
\end{equation}
where
\begin{equation}
S(t)=\frac12\left\langle \hat I(t)\hat I(0)+\hat I(0)\hat I(t)\right\rangle-\left\langle\hat I\right\rangle^2
\label{soft}
\end{equation}
and $\hat I(t)=e^{it\hat H}\hat I e^{-it\hat H}$  is the current operator in the Heisenberg picture, The general relation between
$S(t)$ and $\Pi(\tau)$ can be established through the analytic continuation procedure combined with fluctuation-dissipation theorem.
From the analysis presented in Appendix A one readily finds
\begin{equation}
{\rm Im}\left[\left.\left(\Pi_{i\omega_k}-\beta P\delta_{k,0}\right)\right|_{i\omega_k\to\omega+i0}\right]=\tanh\left(\frac{\omega}{2T}\right)S_\omega ,
\label{cont}
\end{equation}
where
\begin{equation}
P=\langle I^2(\tau)\rangle-\langle I(\tau)\rangle^2,
\label{PP}
\end{equation}
see also Eq. (\ref{PPP}).

It follows immediately from the above results that in the absence of quantum phase slips $P\to 0$ in the zero temperature limit, i.e. PC noise in superconducting rings vanishes identically
at $T \to 0$. On the contrary, at non-zero temperatures PC noise power does not vanish being peaked at zero frequency,
\begin{equation}
S_\omega=2\pi P \delta(\omega),
\end{equation}
where from Eq. (\ref{curcor}) one gets
\begin{multline}
P=e^2T^2\left(\frac{4\lambda v}{\pi L T}+\frac{\vartheta_3^{(2,0)}(\pi\phi_x,e^{-\frac{\pi LT}{2\lambda v}})}{\vartheta_3(\pi\phi_x,e^{-\frac{\pi LT}{2\lambda v}})}\right)\\-e^2T^2\left(\frac{\vartheta_3^{(1,0)}(\pi\phi_x,e^{-\frac{\pi LT}{2\lambda v}})}{\vartheta_3(\pi\phi_x,e^{-\frac{\pi L T}{2\lambda v}})}\right)^2
\label{nzT}
\end{multline}
In the low and high temperature limits this expression reduces to
\begin{equation}
P\approx\begin{cases}
\frac{32e^2v^2\lambda^2}{L^2}e^{-\frac{2\pi v\lambda}{LT}}\cosh\left(\frac{4\pi v\lambda\mathrm{min}_{n}\left(  n+\phi_{x}\right)}{LT}\right) & T\ll \frac{2\pi v\lambda}{L}\\
\frac{4e^2v\lambda T}{\pi L}-8e^2T^2e^{-\frac{\pi LT}{2\lambda v}}\cos(2\pi\phi_x) & T\gg \frac{2\pi v\lambda}{L}
\end{cases}.
\end{equation}

\begin{figure}[h]
\includegraphics[width=0.99\columnwidth]{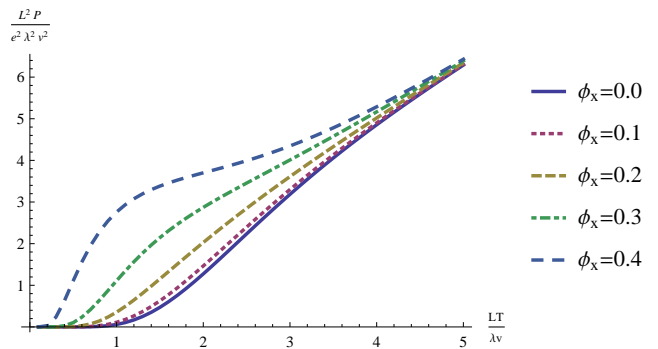}
\caption{(Color online) Temperature dependent zero frequency PC fluctuations in superconducting rings at different values of the magnetic flux $\phi_x$.}
\label{fig2}
\end{figure}
The dependence (\ref{nzT}) is also depicted in Fig. 2 for different values of the magnetic flux. We observe that at sufficiently
low temperatures the magnitude of PC fluctuations can be tuned by the external flux $\phi_x$, hence, indicating coherent nature of such fluctuations. At higher temperatures quantum coherence is destroyed and $P(T) \propto T$ becomes practically independent of $\phi_x$.

\section{Effective action for quantum phase slip rings}

As we already pointed out, the above analysis remains valid only in the case of sufficiently thick rings where quantum fluctuations of the order parameter field can be neglected. In general, the expression
(\ref{partf}) applies only at length and time scales exceeding respectively the superconducting coherence length $\xi \sim \sqrt{D/\Delta}$ and the inverse gap $\Delta^{-1}$. At smaller length and time scales quantum fluctuations of the order parameter field can and do occur in the form of quantum phase slips.

Contributions from these QPS events to both the partition function ${\mathcal Z}$ and the current correlation functions
can be taken into account in the semiclassical approximation
 As usually, in the main approximation it suffices
to take into account all relevant saddle point configurations of the phase variable $\varphi$ which satisfy the equation
\begin{equation}
(\partial_\tau^2+v^2\partial_x^2)\varphi(x,\tau)=0.
\label{speq}
\end{equation}
Apart from trivial solutions of this equation linear in $\tau$ and $x$
there exist nontrivial ones which correspond to virtual phase jumps by $\pm 2\pi$ at various points of a superconducting ring where the magnitude of the order parameter gets locally (at spatial scales $x_0 \sim \xi$) and temporarily (within the time interval $\tau_0 \sim 1/\Delta$) suppressed by quantum fluctuations. These quantum topological objects can be viewed as vortices in space-time and represent QPS events \cite{AGZ,ZGOZ,GZQPS}. 
For sufficiently long wires (rings) and outside
the QPS core $|x| > x_0$, $|\tau | > \tau_0$ (which position in space-time can be chosen, e.g., at $x=0$ and $\tau=0$) the saddle point solution $\tilde{\varphi}(x,\tau )$ corresponding to a single QPS event should satisfy the identity
\begin{equation}
\partial_{x}\partial_{\tau}\tilde{\varphi}-\partial_{\tau}\partial_{x}%
\tilde{\varphi}=2\pi\delta(\tau,x)
\end{equation}
implying that after a wind around the QPS center the phase
should change by $2\pi$. This saddle point solution has the form
\begin{equation}
\tilde{\varphi}(x,\tau)=-\arctan(x/v\tau).\label{QPSphi}
\end{equation}

Analogously one can consider more complicated QPS configurations $\varphi^{qps}(x,\tau)$ consisting of an arbitrary number of phase jump events. The contributions from all these configurations can be effectively summed up with the aid of the approach involving the so-called duality transformation.

In order to proceed let us express the general solution of Eq. (\ref{speq}) in the form
\begin{equation}
  \varphi^{sp}(x,\tau)=a_{m}\tau+b_{n}x+\varphi^{qps}(x,\tau),
\end{equation}
where $a_m$ and $b_n$ are some constants fixed by the boundary conditions. We also introduce the vorticity field $\varpi(x,\tau)$ with the aid of the relations
\begin{equation}
 v\partial_x\varpi=\partial_\tau\varphi^{qps}\quad \partial_\tau\varpi=-v\partial_x\varphi^{qps}.
 \label{dualr}
\end{equation}
This field is single-valued and it obeys the equation
\begin{equation}
\partial^2_\tau\varpi+v^2\partial^2_x\varpi=-2\pi v\sum\limits_j\nu_j\delta(x-x_j)\delta(\tau-\tau_j),
\label{vorteq}
\end{equation}
where $x_j$, $\tau_j$ and $\nu_j$ denote respectively the space and time coordinates of the $j$-th phase slip and as well as its topological charge equal to $\pm 1$ for the phase jump by $\pm 2\pi$. It follows from the path integral boundary conditions (\ref{boucond})  that the vorticity field derivatives are periodical functions in both space and time implying, in turn, that $\sum_j\nu_j=0$.

Let us introduce function
\begin{equation}
\varpi^{qps}(x,\tau)=\frac{\beta Lv}{2\pi }\sum\limits_{|m|+|n|>0}\frac{e^{\frac{2\pi im\tau}{\beta}+\frac{2\pi inx}{L}}}{m^2L^2+n^2v^2\beta^2}.
\end{equation}
It is straightforward to verify that the function
\begin{equation}
\varpi(x,\tau)=\sum\limits_j\nu_j\varpi^{qps}(x-x_j,\tau-\tau_j)
\end{equation}
satisfies Eq. (\ref{vorteq}) and by virtue of the duality relations (\ref{dualr}) it yields the saddle point configuration $\varphi^{qps}$. Combining the boundary conditions (\ref{boucond}) with the above equations, one easily finds
\begin{multline}
\varphi^{qps}(L,\tau)-\varphi^{qps}(0,\tau)=-\frac{1}{v}\int\limits_0^Ldx\partial_\tau\varpi(x,\tau)\\
=2\pi\sum\limits_j\nu_j\left(\theta(\tau-\tau_j)+\frac{\tau_j}{\beta}\right),
\end{multline}
\begin{multline}
\varphi^{qps}(x,\beta)-\varphi^{qps}(x,0)=v\int\limits_0^\beta d\tau\partial_x\varpi(x,\tau)\\
=-2\pi\sum\limits_j\nu_j\left(\theta(x-x_j)+\frac{x_j}{L}\right)
\end{multline}
and, hence,
\begin{equation}
a_m=\frac{2\pi}{\beta}\left(m+\sum\limits_j\nu_j\frac{x_j}{L}\right),
\end{equation}
\begin{equation}
 b_n=\frac{2\pi}{L}\left(n+\phi_x
-\sum\limits_j\nu_j\frac{\tau_j}{\beta}\right).
\end{equation}
These equations complete our analysis of relevant saddle point configurations. For each of these configurations
the expression for the current (\ref{current0}) takes the form
\begin{equation}
I(\tau)=\frac{4 ev\lambda}{L}\left(n+\phi_x+\sum\limits_j\nu_j\theta(\tau-\tau_j)\right).
\end{equation}

Let us now carry out the summation over all possible saddle point configurations. Making use of the results \cite{ZGOZ,GZQPS}, for the
partition function we get
\begin{widetext}
\begin{multline}
\mathcal Z[J(\tau)]=\sum\limits_{N=0}^\infty\frac{1}{N!}\sum\limits_{\nu_1,..,\nu_N=\pm1}\delta_{\sum_j\nu_j,0}\int dx_1d\tau_1...dx_Nd\tau_N\sum\limits_{m,n=-\infty}^\infty e^{-\frac{\lambda}{2\pi}\left(\frac{\beta L}{v}a_m^2+\beta L v b_n^2\right)}\\
\times\left(\frac{\gamma_{QPS}}{2}\right)^N e^{-\frac{\lambda}{2\pi}\int dxd\tau\left(v(\partial_x\varpi)^2+v^{-1}(\partial_\tau\varpi)^2\right)+i\int d\tau J(\tau)I(\tau)}.
\label{pfqps1}
\end{multline}
Here $J(\tau)$ is the source variable introduced for future calculation of the current correlation functions and $\gamma_{QPS}$ is the QPS rate per unit wire length defined as \cite{GZQPS}
\begin{equation}
 \gamma_{QPS} \sim (g_\xi \Delta /\xi )\exp (-a g_\xi )
\label{gqps}
\end{equation}
where $g_\xi = 4\pi N_0Ds/\xi $ is the dimensionless conductance of the wire segment of length equal to the
superconducting coherence length $\xi$ and $a$ is an unimportant numerical prefactor of order one. Eqs. (\ref{pfqps1}), (\ref{gqps}) apply
provided $g_\xi$ remains sufficiently large, i.e. $g_\xi e^{-ag_\xi /2}\ll 1$.

Rewriting the sum over $m,n$ by means of the relations
\begin{equation}
\sum\limits_{m=-\infty}^\infty e^{-\frac{\lambda\beta L}{2\pi v}a_m^2}\sim\sum\limits_{m=-\infty}^\infty e^{-\frac{\pi v\beta m^2}{2\lambda L}+2\pi i m\sum\limits_j\nu_j\frac{x_j}{L}},
\end{equation}
\begin{equation}
\sum\limits_{n=-\infty}^\infty e^{-\frac{\lambda\beta vL}{2\pi }b_n^2+\frac{4 i ev\lambda(n+\phi_x)}{L}\int d\tau J(\tau)}\sim\sum\limits_{n=-\infty}^\infty e^{-\frac{\pi L }{2\lambda \beta v}\left(n-\frac{2ev\lambda}{\pi L}\int d\tau J(\tau)\right)^2+2\pi in\phi_x -2\pi i \left(n-\frac{2ev\lambda}{\pi L}\int d\tau J(\tau)\right)\sum\limits_j\nu_j\frac{\tau_j}{\beta}},
\end{equation}
employing the Kronecker delta-function representation $\delta_{m,n}=\int_0^{2\pi} dz e^{iz(m-n)}/(2\pi)$
and formally inserting the path integral over the $\varpi$-field,
we obtain
\begin{multline}
\mathcal Z[J(\tau)]\sim\sum\limits_{N=0}^\infty\frac{1}{N!}\sum\limits_{\nu_1,..,\nu_N=\pm1}\int\limits_0^{2\pi}\frac{dz}{2\pi}\int dx_1d\tau_1...dx_Nd\tau_N\sum\limits_{m,n=-\infty}^\infty e^{2\pi in\phi_x-\frac{\pi v\beta m^2}{2\lambda L}-\frac{\pi L }{2\lambda \beta v}\left(n-\frac{2ev\lambda}{\pi L}\int d\tau J(\tau)\right)^2}\\
\times\left(\frac{\gamma_{QPS}}{2}\right)^N e^{2\pi i m\sum\limits_j\nu_j\frac{x_j}{L} -2\pi i \left(n-\frac{2ev\lambda}{\pi L}\int d\tau J(\tau)\right)\sum\limits_j\nu_j\frac{\tau_j}{\beta}}
\int\mathcal D\varpi
e^{-\frac{\lambda}{2\pi}\int dxd\tau\left(v(\partial_x\varpi)^2+v^{-1}(\partial_\tau\varpi)^2\right)}\\
\times e^{iz\sum\limits_j\nu_j+\frac{4iev\lambda}{L}\sum\limits_j\nu_j\int d\tau J(\tau)\theta(\tau-\tau_j)}\delta\left(\partial^2_\tau\varpi+v^2\partial^2_x\varpi+2\pi v\sum\limits_j\nu_j\delta(x-x_j)\delta(\tau-\tau_j)\right),
\label{funcdf}
\end{multline}
where the functional delta-function follows from Eq. (\ref{vorteq}). Expressing this delta-function via the integral over the dual field $\eta(x,\tau)$ with the periodic boundary conditions and performing gaussian integration over $\varpi$, from Eq. (\ref{funcdf}) we get
\begin{multline}
\mathcal Z[J(\tau)]\sim\sum\limits_{N=0}^\infty\frac{1}{N!}\sum\limits_{\nu_1,..,\nu_N=\pm1}\int\limits_0^{2\pi}\frac{dz}{2\pi}\int dx_1d\tau_1...dx_Nd\tau_N\sum\limits_{m,n=-\infty}^\infty e^{2\pi in\phi_x-\frac{\pi v\beta m^2}{2\lambda L}-\frac{\pi L }{2\lambda \beta v}\left(n-\frac{2ev\lambda}{\pi L}\int d\tau J(\tau)\right)^2}\\
\times\left(\frac{\gamma_{QPS}}{2}\right)^N e^{2\pi i m\sum\limits_j\nu_j\frac{x_j}{L} -2\pi i \left(n-\frac{2ev\lambda}{\pi L}\int d\tau J(\tau)\right)\sum\limits_j\nu_j\frac{\tau_j}{\beta}}
\int\mathcal D\eta
e^{-\frac{\pi v}{2\lambda}\int dxd\tau\left((\partial_\tau\eta)^2+v^2(\partial_x\eta)^2\right)}\\
\times e^{2\pi i v\sum\limits_j\nu_j\eta(x_j,\tau_j)+iz\sum\limits_j\nu_j+\frac{4iev\lambda}{L}\sum\limits_j\nu_j\int d\tau J(\tau)\theta(\tau-\tau_j)}.
\end{multline}
Finally, introducing a new field
\begin{equation}
\chi(x,\tau)=-\frac{2\pi m x}{L}+\frac{2\pi\tau}{\beta}\left(n-\frac{2ev\lambda}{\pi L}\int d\tau J(\tau)\right)-2\pi v\eta(x,\tau)-z-\frac{4ev\lambda}{L}\int d\tau' J(\tau')\theta(\tau'-\tau)
\label{chi}
\end{equation}
\end{widetext}
obeying the boundary conditions $\chi(x,\beta)-\chi(x,0)=2\pi n$, $\chi(L,\tau)-\chi(0,\tau)=-2\pi m$, we arrive at the resulting expression for the generating functional
\begin{equation}
\mathcal Z[J(\tau)]\sim\sum\limits_{m,n=-\infty}^\infty e^{2\pi i n\phi_x}\int^{mn} \mathcal D\chi e^{-S_{\rm eff}[\chi(x,\tau),J(\tau)]}
\label{ZJ}
\end{equation}
with the effective action
\begin{multline}
S_{\rm eff}=
\frac{1}{8\pi\lambda v}\int\limits_0^\beta d\tau\int\limits_0^L dx[(\partial_\tau\chi(x,\tau)-4ev\lambda J(\tau)/L)^2\\+v^2(\partial_x\chi(x,\tau))^2]
-\gamma_{QPS}\int\limits_0^\beta d\tau\int\limits_0^Ldx\cos(\chi(x,\tau)).
\label{efacchi}
\end{multline}
These expressions conclude our derivation of the generating functional
and the effective action for superconducting nanorings with quantum phase slips. We observe that the original problem of QPS rings pierced by the magnetic flux can be exactly mapped onto a sine-Gordon model on torus. Eqs. (\ref{ZJ}), (\ref{efacchi}) fully account for interactions between different QPS and serve as a convenient starting point for further analysis of the ground state properties of QPS rings.

As we already pointed out, in the absence of the source $J(\tau) \to 0$ the effective action (\ref{efacchi}) turns out to be exactly dual to that for spatially extended quasi-one-dimensional Josephson barriers \cite{BP}. Namely, the Josephson phase $\varphi (x,\tau)$ in the latter model is dual to the field $\chi (x,\tau)$ (\ref{chi}) which defines the electric charge $q(x,\tau)=\chi (x,\tau)/2\pi$ passing in the superconducting wire through the point $x$ at the time moment $\tau$.
Accordingly, one model is mapped onto the other if we interchange
the flux quantum and the Cooper pair charge $\Phi_0 \leftrightarrow 2e$, the Swihart velocity \cite{Swi} and the Mooij-Sch\"on velocity $v$, the Josephson penetration length $\lambda_J$ (which controls the size of fluxons) and the parameter $\lambda_{2e}=\sqrt{v/(4\pi \lambda \gamma_{QPS})}$ (which defines the size of $2e$ charge solitons in QPS junctions) and so on.
All the phenomena known to exist in long Josephson junctions will have their dual analogues in QPS junctions and rings.

At low enough temperatures $T \ll v/L$  and provided the ring perimeter $L=2\pi R$ remains sufficiently small (see below)
one can ignore the spatial dependence of the field $\chi$ and, hence, neglect the term $v^2(\partial_x\chi)^2$ in the effective action (\ref{efacchi}). Then our problem reduces to a zero-dimensional one with an effective Hamiltonian \cite{SZ10,SZ11}
\begin{equation}
   \hat H=\frac{E_R}{2}(\hat \phi -\phi_x)^2+U_0 (1-\cos \chi )
\label{H1}
\end{equation}
describing a fictitious quantum particle on a ring in the presence of the cosine external potential.  Here we identify \cite{SZ12}
\begin{equation}
E_R=\frac{\pi^2 N_0 D\Delta s}{R}\sim \frac{g_{\xi}\Delta\xi}{R}
\label{ER}
\end{equation}
and
\begin{equation}
U_0=2\pi R\gamma_{QPS} \sim \frac{g_{\xi}\Delta R}{\xi} e^{-ag_\xi}.
\label{U0}
\end{equation}

\section{Persistent current and noise in QPS rings}

Let us now employ the above formalism in order to evaluate both the average value of PC $I$ and the current-current correlation function $\Pi(\tau )$ in quantum phase slip rings. Taking the derivatives of the generating functional $\mathcal Z[J(\tau)]$ over the source variable $J(\tau)$ and setting this variable equal to zero afterwards, we express both these quantities in terms of the $\chi$-field as
\begin{equation}
I\equiv\langle I(\tau)\rangle =-\frac{i e}{\pi}\langle\partial_\tau\chi(x,\tau)\rangle
\end{equation}
\begin{multline}
\Pi(\tau_1-\tau_2)=\frac{4e^2\lambda v}{\pi L}\delta(\tau_1-\tau_2)\\-\frac{e^2}{\pi^2L^2}\int dx_1 dx_2\langle
\partial_{\tau_1}\chi(x_1,\tau_1)\partial_{\tau_2}\chi(x_2,\tau_2)\rangle-I^2
\label{pi1}
\end{multline}
One can also decompose the source variable as $J(\tau)=J_0+\partial_\tau J_1(\tau)$ and perform a shift under the functional integral
\begin{multline}
\mathcal Z[J_0+\partial_\tau J_1(\tau)]=\sum_{mn}e^{2\pi i n\phi_x+2eJ_0n-\frac{2e^2v\lambda J_0^2\beta}{\pi L}}\\
\times\int^{mn} \mathcal D\chi e^{-S_{\rm eff}[\chi(x,\tau)+4ev\lambda J_1(\tau)/L,0]}
\end{multline}
Expanding both sides in powers of $J_0$ and $J_1(\tau)$ we reproduce
the standard expression for the current
\begin{equation}
I=-\frac{e}{\pi\beta}\frac{\partial\ln\mathcal Z}{\partial\phi_x}
\label{PCav}
\end{equation}
and also arrive at the following exact relations
\begin{equation}
\int d\tau \Pi(\tau)=\frac{4e^2\lambda v }{\pi L}\left(1+\frac{L}{4\pi \lambda\beta v}\frac{\partial^2\ln\mathcal Z}{\partial\phi_x^2}\right),
\end{equation}
\begin{multline}
\Pi''(\tau)=-\frac{16 \gamma_{QPS}e^2\lambda^2v^2}{L^2}\int dx\langle\cos(\chi(x,\tau_1))\rangle\delta(\tau)\\+\frac{16 \gamma_{QPS}^2e^2\lambda^2v^2}{L^2}\int dx_1dx_2\langle\sin(\chi(x_1,\tau))\sin(\chi(x_2,0)) \rangle .
\label{pi2}
\end{multline}

Consider first the average value of PC (\ref{PCav}). In the zero dimensional limit described by the effective Hamiltonian (\ref{H1})
the result is well known. At low temperatures $T \to 0$ Eq. (\ref{PCav}) reduces to
\begin{equation}
I=\frac{e}{\pi}\frac{\partial \epsilon_0(\phi_{x})}{\partial\phi_{x}},
\end{equation}
where $\epsilon_0(\phi_{x})$ is the flux-dependent ground state energy. For smaller rings with $U_0\ll E_{R}$ one has
\begin{equation}
\epsilon_{0}(\phi_x)=\frac{E_{R}}{2\pi^{2}}\arcsin^{2}\left[\left( 1-\frac{\pi^{2}
}{2}\left(\frac{U_0}{E_{R}}\right)^{2}\right)\sin(\pi\phi_{x})\right]
,
\end{equation}
i.e. the ground state energy is almost parabolic except in the vicinity of the crossing points $\phi_x=1/2+n$ where the gap to the first excited energy band $\delta E_{01}=U_0$  opens up due to level repulsion. Accordingly, not too close to the points $\phi_x=1/2+n$
PC is not sensitive to quantum phase slips and is again defined by Eq.
(\ref{evcur}).

For larger $U_0$ the bandwidth shrinks while the gaps become bigger. In the limit $U_0 \gg E_R$ PC changes its dependence on $\phi_x$ from the sawtooth one (\ref{evcur}) to \cite{AGZ,Z10,MLG}
\begin{equation}
I=I_{0}\sin(2\pi\phi_{x}).
\label{tok}
\end{equation}
In this limit, i.e. for $R\gg R_c \sim \xi \exp (ag_\xi /2)$, we find \cite{Z10}
\begin{equation}
I_0\sim E_R^{1/4} U_0^{3/4} e^{-R/R_{c}}
\label{lsc1}
\end{equation}
with $E_R$ and $U_0$ defined respectively in Eqs. (\ref{ER}) and (\ref{U0}).

Let us now include the effect of logarithmic inter-QPS interactions which is accounted for by the sine-Gordon effective action (\ref{efacchi}). This task can be accomplished with the aid of
the well known Berezinskii-Kosterlitz-Thouless (BKT) renormalization group (RG) approach. The corresponding RG equations read
\begin{equation}
\frac{d\zeta}{d\ln \Lambda}=(2-\lambda)\zeta\qquad \frac{d\lambda}{d\ln\Lambda}=-32\pi^2\zeta^2 \lambda^2 K(\lambda),
\label{BKT}
\end{equation}
where $\zeta=\gamma_{QPS}\Lambda^2$ is the dimensionless coupling parameter, $\Lambda$ is the renormalization scale and $K(\lambda)$ is some nonuniversal function (which depends on the renormalization scheme) equal to one at the quantum BKT phase transition point $\lambda=2$ which separates and superconducting (ordered) phase $\lambda > 2$ with bound QPS-antiQPS pairs and a disordered phase $\lambda < 2$ with unbound QPS \cite{ZGOZ}.

As usually, we start renormalization at the shortest scale $\Lambda \sim \xi_c
=\sqrt{\xi^2+v^2/\Delta^2}$ and proceed to bigger scales. Within the first order  perturbation theory in $\zeta$ one can ignore weak renormalization of the parameter $\lambda$. Within this accuracy the solution of Eqs. (\ref{BKT}) takes the simple form $\gamma_{QPS}(\Lambda)=\gamma_{QPS}(\xi_c/\Lambda)^\lambda$. Our RG procedure should be stopped at the maximum scale equal to the ring perimeter $\Lambda \sim L=2\pi R$. In this way we arrive at the renormalized QPS rate for our system
\begin{equation}
\tilde \gamma_{QPS}=\gamma_{QPS}(\xi_c/L)^\lambda .
\label{ren}
\end{equation}
This result demonstrates that interaction-induced renormalization effects remain weak and can be disregarded only for very small values of $\lambda \ll 1/\ln (L/\xi_c )$.

Substituting the renormalized rate (\ref{ren}) instead of the bare one into Eq.
(\ref{U0}), we again reproduce the same expressions for PC, now with $\gamma_{QPS} \to \tilde \gamma_{QPS}$. As before, for smaller rings
with $R \ll \tilde R_c$ and at $T \to 0$ the current is defined
by Eq. (\ref{evcur}) at all values of the flux except for $\phi_x \approx 1/2+n$ where QPS effects with the effective rate (\ref{ren})
become significant. The critical radius $\tilde R_c$ is now
determined by the condition $E_R \sim 2\pi \tilde\gamma_{QPS}$ which yields
\begin{equation}
\tilde R_c \sim \xi \exp \left(\frac{ag_\xi}{2-\lambda}\right)
\left(\frac{\xi}{\xi_c}\right)^{\frac{\lambda}{2-\lambda}},
\label{rc}
\end{equation}
where $\lambda$ is supposed not to exceed 2.
In the opposite limit of bigger rings $R \gg \tilde R_c$ we again
reproduce Eq. (\ref{tok}), where now
\begin{equation}
I_0 \sim \kappa^{3/4} g_{\xi}\Delta\left(\frac{R}{\xi}\right)^{1/2-3\lambda /4} e^{-3ag_\xi /4-R/\tilde R_{c}},
\label{lsc2}
\end{equation}
where $\kappa = (\xi_c/\xi )^{\lambda}$.

From Eq. (\ref{rc}) we observe that the critical radius $\tilde R_{c}$
(beyond which PC gets exponentially suppressed) increases with increasing $\lambda$ and eventually diverges at the
quantum BKT phase transition point $\lambda \to 2$. In the ordered phase $\lambda > 2$ QPS become practically irrelevant and PC is defined by Eq. (\ref{evcur}) for any value of $R$.

Now let us analyze fluctuations of PC in QPS rings. In what follows we will restrict our attention to the most interesting low temperature limit $T \to 0$. We first evaluate the imaginary time current-current correlator  (\ref{pi}), (\ref{pi1}) and then perform its analytic continuation to real times. It will be convenient for us to make use of Eq. (\ref{pi2}) which can be evaluated perturbatively in $\gamma_{QPS}$ provided $R \ll \tilde R_c$. The whole procedure is described in details in Appendix B. It yields the result
\begin{widetext}
\begin{equation}
\Pi_{i\omega}=\frac{8\gamma_{QPS}^2 e^2 \lambda^2 v^2  }{\Gamma^2(\lambda)}\left(\frac{8\pi^2\xi_c^2}{L^2}\right)^\lambda
\sum\limits_{k=0}^\infty\frac{\Gamma^2(\lambda+k)}{\Gamma^2(1+k)}
\left(\frac{1}{\Omega_k(\phi_x)(\omega^2+\Omega_{k}^2(\phi_x))}
+\frac{1}{\Omega_k(-\phi_x)(\omega^2+\Omega_k^2(-\phi_x))}\right),
\label{piom}
\end{equation}
\end{widetext}
where
\begin{equation}
\Omega_k(\phi_x)=\frac{2\pi \lambda v(1+2\phi_x)}{L}+\frac{4\pi v k}{L}
\label{diff}
\end{equation}
denote the energy differences between the exited states and the ground state of our ring. Eq. (\ref{diff}) applies for $-1/2<\phi_x \leq 1/2$. Outside this interval $\Omega_k(\phi_x)$ should be continued periodically with the period equal to unity.

The result (\ref{piom}) is well suited for our analytic continuation to real time which can now be performed in a straightforward manner with the aid of Eq. (\ref{cont}). As a result, we arrive at the final expression for PC noise power spectrum at $T=0$:
\begin{multline}
S_\omega=\frac{4\pi\gamma_{QPS}^2 e^2 \lambda^2 v^2  }{\Gamma^2(\lambda)}\left(\frac{8\pi^2\xi_c^2}{L^2}\right)^\lambda
\sum\limits_{k=0}^\infty\frac{\Gamma^2(\lambda+k)}{\Gamma^2(1+k)}
\\\times\left(\frac{1}{\Omega_{k}^2(\phi_x)}
\delta(\omega-\Omega_k(\phi_x))+\frac{1}{\Omega_{k}^2(\phi_x)}
\delta(\omega+\Omega_k(\phi_x))\right.\\
+\frac{1}{\Omega_{k}^2(-\phi_x)}\delta(\omega-\Omega_k(-\phi_x))\\
\left.+\frac{1}{\Omega_{k}^2(-\phi_x)}
\delta(\omega+\Omega_k(-\phi_x))\right).
\label{final}
\end{multline}
We observe that in accordance with general considerations \cite{SZ10,SZ11} $S_\omega$ depends periodically on the external magnetic flux $\phi_x$ and consists of sharp peaks at the
frequencies $\Omega_k$ corresponding to the system eigenmodes.
These features clearly illustrate coherent nature of PC noise.
\begin{figure}[h]
a)\includegraphics[width=0.44\columnwidth]{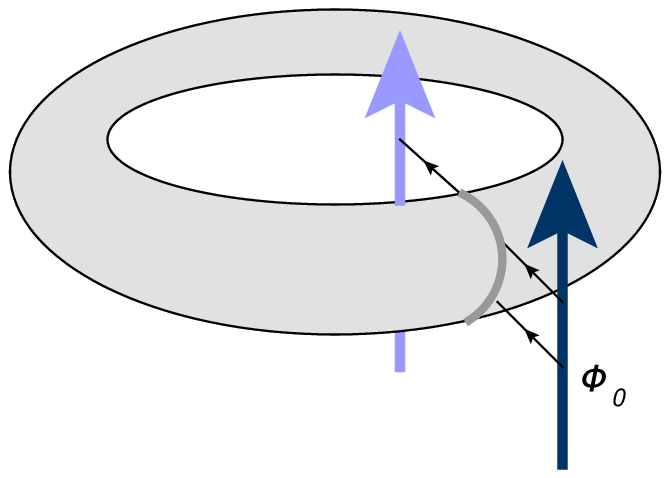}
b)\includegraphics[width=0.44\columnwidth]{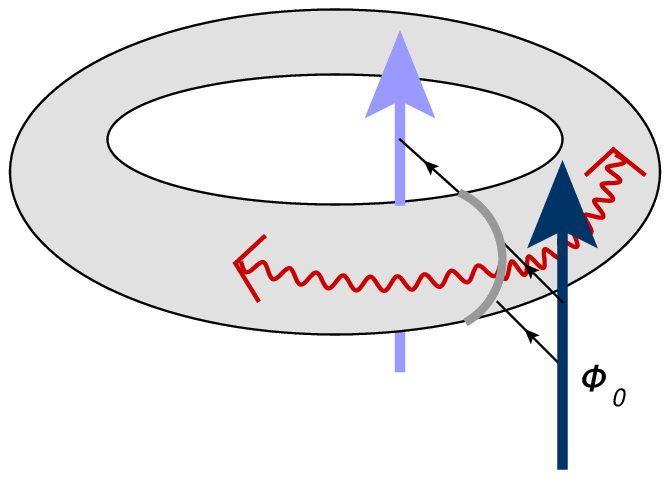}
\caption{(Color online) The process of coherent flux tunneling (a) without
and (b) with excitation of a pair of Mooji-Sch\"on plasma modes.}
\label{fig3}
\end{figure}
The effect of inter-QPS interactions on PC noise turns out to be
somewhat richer than that for the average value of PC. On one hand,
interactions-induced renormalization of the QPS rate (\ref{ren})
shows up here as well: The magnitude of PC noise power (\ref{final}) scales with $\tilde \gamma_{QPS}^2 \propto L^{2\lambda -2}$.
On the other hand, apart from this renormalization effect
interactions also yield extra peaks in PC noise as compared
to the case of small rings described by the Hamiltonian (\ref{H1}). Indeed, the terms with $k=0$ in Eq. (\ref{final}) originating from virtual transitions of the flux quanta both into and out of the ring
(see also Fig. \ref{fig3}a) are fully analogous to those already established in the absence of QPS interactions \cite{SZ12}. At the same time, all extra terms
with $k\neq 0$ correspond to virtual flux quanta transitions accompanied by creation of Mooji-Sch\"on plasmon excitations.
Such contributions to PC noise are absent in the zero dimensional limit \cite{SZ12}. Keeping in mind the momentum conservation one can conclude that the simplest process of that kind implies simultaneous creation of two plasmon excitations with opposite momenta values propagating clockwise and counterclockwise around the ring. Such processes are also illustrated in Fig. \ref{fig3}b.

Let us emphasize again that -- owing to its coherent nature -- PC noise in superconducting nanorings can be tuned by the external magnetic flux. Both the positions of the peaks and the magnitude of this noise essentially depend on $\phi_x$. As follows from Eq. (\ref{final}), in the immediate vicinity of level degeneracy points $\phi_x =\pm 1/2$ fluctuations of PC become strong and our perturbative in $\gamma_{QPS}$ analysis becomes inapplicable even for $R \ll \tilde R_c$. In this case it is necessary to account for level splitting
and regularize the corresponding terms in Eq. (\ref{final}). Roughly speaking,
this regularization amounts to substituting the value $\sim U_0$ instead of $\Omega_0(\phi_x)$ whenever the former exceeds the latter.

\section{Discussion}
In this work we investigated the effect of interacting quantum phase slips on persistent current and its fluctuations in ultrathin superconducting rings and nanowires forming the so-called QPS junctions (Fig. \ref{fig1}).

Starting from the low-energy effective action \cite{ZGOZ} which describes dynamics of the superconducting phase in such nanowires we evaluated the partition function of the problem and mapped the original problem onto the effective sine-Gordon theory. In this way we emphasized a complete duality between our problem and that of long Josephson junctions studied in the literature over last decades.

We analyzed the effect of electrodynamical interactions between
quantum phase slips and demonstrated that the average value of PC
in QPS rings can be well described with the aid of the effective
Hamiltonian (\ref{H1}) for a fictitious quantum particle moving on a ring in
a periodic potential with the height $U_0$ proportional to the renormalized QPS rate (\ref{ren}), i.e
\begin{equation}
\frac{U_0}{\Delta} \sim \kappa g_{\xi}\left(\frac{L}{\xi}\right)^{1-\lambda} e^{-ag_\xi}.
\label{U0r}
\end{equation}
Note that the value $U_0$ was directly measured in recent experiments
\cite{Ast13} for a number of samples with different cross section values $s$. Comparing these data with Eq. (\ref{U0r}) it is necessary
to bear in mind that $g_\xi \propto s$ and $\lambda \propto \sqrt{s}$.
For the sample parameters \cite{Ast13} and estimating the capacitance per unit length similarly to \cite{ZGOZ} as $C\sim 1\div 2$ we find $g_\xi\approx 90$, $v\sim 7\times10^5 {\rm m/s}$ and $\lambda\sim 0.5$. Combining these numbers with the data points
for $U_0$ \cite{Ast13} we also estimate $a \sim 0.14$. It is easy to observe that for such system parameters QPS interaction effects can be important and in general need to be accounted for in order to provide a quantitative comparison with experiment.

In contrast to the average PC value, fluctuations of PC cannot in general be adequately described by the Hamiltonian (\ref{H1}) even if
the renormalization of $\gamma_{QPS}$ (\ref{ren}) is taken into  account. This is because virtual tunneling of flux quanta across the
superconducting wire in general leads to creation of plasmon modes in the system (see Fig. \ref{fig3}b) thereby causing extra peaks in the PC noise power spectrum (\ref{final}). Thus, by experimentally detecting these peaks one can directly demonstrate the existence of Mooij-Sch\"on plasma modes in superconducting nanorings \cite{FN}.

An important feature of low temperature PC noise is that it can be tuned by an externally applied magnetic flux. This feature clearly illustrates coherent nature of persistent current noise in QPS rings. Here we evaluated the dependence $S_\omega (\phi_x)$ in the experimentally relevant limit $R < \tilde R_c$ in which case one can
proceed perturbatively in the QPS rate $\gamma_{QPS}$. In the opposite
limit $R>\tilde R_c$ PC noise also has the form of sharp peaks although its dependence on the magnetic flux becomes much weaker \cite{SZ12}.
At non-zero temperatures PC noise is modified in two ways: (i) a zero frequency peak (\ref{nzT}) appears which is not related to QPS and (ii) numerous extra QPS-related peaks at non-zero
frequencies emerge, cf. Eq. (\ref{soft2}). At low enough $T$ quantum coherence is still maintained, however with increasing temperature
the dependence on $\phi_x$ gets less pronounced and PC noise eventually becomes incoherent.

Finally let us point out a certain physical similarity between
PC noise studied here and the equilibrium supercurrent in point contacts between superconductors \cite{AverinI,Madrid,GZ10}. Also in the latter case the noise power spectrum depends on the phase difference across the superconducting weak link and has the form of peaks which occur both at zero and non-zero frequencies. Similarly to our problem, at $T\to 0$ the zero frequency peak
disappears while the other peaks do not vanish except in the limit
of fully transparent barriers \cite{AverinI,Madrid,GZ10}. Unlike here, however, in the case of superconducting contacts the noise peaks at non-zero frequencies have to do with subgap Andreev levels inside such contacts and are not related to quantum phase slips.

\vspace{0.5cm}

\centerline{\bf Acknowledgement}

This work was supported in part by RFBR Grant No. 12-02-00520-a.

\appendix
\section{}

The quantities $\Pi(\tau)$ and $S(t)$ defined respectively in Eqs. (\ref{pi}) and (\ref{soft}) can be related to each other through the appropriate analytic continuation procedure combined with the fluctuation-dissipation theorem. Expressing both correlators $\Pi(\tau)$ and $S(t)$ in terms of the exact eigenstates $\epsilon_m$ of the system Hamiltonian $\hat H|m\rangle=\epsilon_m|m\rangle$, we find
\begin{equation}
 \Pi_{i\omega_k}=\beta P\delta_{k,0}+\frac{1}{\mathcal Z}\sum\limits_{m\neq n}|\langle m|\hat I|n\rangle|^2\frac{e^{-\beta \epsilon_m}-e^{-\beta \epsilon_n}}{i\omega_k+\epsilon_n-\epsilon_m}
 \label{pi0}
\end{equation}
\begin{multline}
S_\omega=2\pi P\delta(\omega)+\frac{\pi}{\mathcal Z}\sum\limits_{m\neq n}|\langle m|\hat I|n\rangle|^2\left(e^{-\beta \epsilon_n}+e^{-\beta \epsilon_m}\right)\\\times\delta(\omega+\epsilon_n-\epsilon_m)
\label{soft2}
\end{multline}
where
\begin{equation}
P=\frac{1}{\mathcal Z}\sum\limits_n|\langle n|\hat I|n\rangle|^2e^{-\beta \epsilon_n}-I^2
\label{PPP}
\end{equation}
defines the zero-frequency contribution, $\mathcal Z=\sum_ne^{-\beta \epsilon_n}$ is the grand partition function and $I=\mathcal Z^{-1}\sum_n\langle n|\hat I|n\rangle e^{-\beta \epsilon_n}$ is the expectation value for the current. With the aid of the above general expressions one easily arrives at Eq. (\ref{cont}) which enables one to recover the current noise power spectrum $S_\omega$ from the imaginary time analysis.

It follows from Eqs. (\ref{soft2}), (\ref{PPP}) that in the zero temperature limit (i) $P\equiv0$, i.e. zero frequency PC noise vanishes identically and (ii) at non-zero frequencies PC noise also vanishes  provided the current operator commutes with the system Hamiltonian $\hat H$. Otherwise at $T \to 0$ fluctuations of PC occur due to virtual transitions between the ground state and the excited states with non-zero matrix elements of the current operator.
In the true zero temperature limit, i.e. at temperatures well below the energy difference between the first excited state and the ground state
Eqs. (\ref{pi0}) and (\ref{soft2}) reduce to
\begin{equation}
\Pi_{i\omega}(T\to0)=2\sum\limits_{m\neq 0}|\langle m|\hat I|0\rangle|^2\frac{\epsilon_m-\epsilon_0}{\omega^2+(\epsilon_m-\epsilon_0)^2}
\end{equation}
\begin{multline}
S_\omega(T\to0)=\pi\sum\limits_{m\neq 0}|\langle m|\hat I|0\rangle|^2(\delta(\omega+\epsilon_0-\epsilon_m)
\\+\delta(\omega+\epsilon_m-\epsilon_0)).
\end{multline}

\section{}
Proceeding perturbatively in $\gamma_{QPS}$, in the leading approximation one can reduce Eq. (\ref{pi2}) to the form
\begin{multline}
\Pi''(\tau)=\frac{8\gamma_{QPS}^2e^2\lambda^2v^2}{L}\int dx\langle\cos(\chi(x,\tau)-\chi(0,0))\rangle_0\\
-\frac{8\gamma_{QPS}^2e^2\lambda^2v^2}{L}\delta(\tau)\int d\tau_1 dx\langle\cos(\chi(x,\tau_1)-\chi(0,0))\rangle_0,
\label{ccni}
\end{multline}
where averaging $\langle ...\rangle_0$ is now performed with the
effective action in the non-interacting limit $\gamma_{QPS}=0$.

The task at hand is to evaluate the correlation function
\begin{multline}
\langle e^{i(\chi(x,\tau)-\chi(0,0))} \rangle_0\\=\sum\limits_{m,n=-\infty}^\infty e^{2\pi in\phi_x} \int^{mn}\mathcal D\chi
e^{i(\chi(x,\tau)-\chi(0,0))}
 \\\times e^{-\frac{1}{8\pi\lambda v}\int\limits_0^\beta d\tau\int\limits_0^L dx\left((\partial_\tau\chi)^2+v^2(\partial_x\chi)^2\right)}.
\label{niav}
\end{multline}
For this purpose it is useful to rewrite this correlation function through the zero topological sector $m=n=0$ as
\begin{multline}
\langle e^{i(\chi(x,\tau)-\chi(0,0))} \rangle_0=\frac{1}{\mathcal Z_0}\frac{\int^{00}\mathcal D\chi e^{-S_0+i(\chi(x,\tau)-\chi(0,0))}}{\int^{00}\mathcal D\chi e^{-S_0}}\\
\times\sum\limits_{mn} e^{-\frac{2\pi gv\beta(\phi_x+m+\tau/\beta)^2}{L}-\frac{\pi\beta v n^2}{2gL}+\frac{2\pi i n x}{L}},
\label{B2}
\end{multline}
where
\begin{equation}
S_0= \frac{1}{8\pi\lambda v}\int\limits_0^\beta d\tau\int\limits_0^L dx\left((\partial_\tau\chi)^2+v^2(\partial_x\chi)^2\right)
\end{equation}
is the noninteracting effective action and
\begin{equation}
\mathcal Z_0=\sum_{mn}e^{-\frac{2\pi\lambda v\beta(m+\phi_x)^2}{L}-\frac{\pi\beta v n^2}{2\lambda L}}.
\end{equation}
Performing gaussian integration in Eq. (\ref{B2}), we obtain
\begin{multline}
 \langle e^{i(\chi(x,\tau)-\chi(0,0))} \rangle_0=\frac{e^{G(x,\tau)-G(0,0)}}{\mathcal Z_0}\\\times\sum\limits_{mn} e^{-\frac{2\pi gv\beta(\phi_x+m+\tau/\beta)^2}{L}-\frac{\pi\beta v n^2}{2gL}+\frac{2\pi i n x}{L}},
\end{multline}
where $G(x,\tau)$ is the Green function in the noninteracting theory (with subtracted zero mode) obeying the equation
\begin{equation}
\left(-\partial_{\tau}^2-v^2\partial_{x}^2\right)G(x,\tau)=4\pi \lambda v(\delta(\tau)\delta(x)-1)
\end{equation}
with periodic boundary conditions. This equation can be resolved with the aid of the Fourier transformation which yields
\begin{multline}
G(x,\tau)-G(0,0)=\frac{2\pi \lambda v \tau^2}{\beta L}\\-\lambda\sum\limits_{m=-\infty}^\infty\ln\left(\frac{\cosh(2\pi v(\tau+\beta m)/L  )-\cos(2\pi x/L)}{\cosh(2\pi v(\beta m)/L  )-1}\right)
\end{multline}
The divergent term with $m=0$ in this sum is regularized bearing in mind that the above expressions do not apply at the space and time scales $\lesssim x_0,\tau_0$, where $x_0\sim \xi$ and $\tau_0 \sim 1/\Delta$. Hence, one can make a replacement $G(0,0)\to G(x_0,\tau_0)$. As a result we get the term with $m=0$ proportional to $$\sim\ln\left(\frac{\cosh(2\pi v\tau/L  )-\cos(2\pi x/L)}{4\pi^2\xi_c^2/L^2}\right).$$

In the interesting for us zero temperature limit the above equations yield
\begin{multline}
\langle\cos(\chi(x,\tau)-\chi(0,0))\rangle_0\\= \left(\frac{4\pi^2\xi_c^2}{L^2}\right)^\lambda\frac{\cosh(4\pi \lambda v\phi_x\tau/L)}{(\cosh(2\pi v\tau/L)-\cos(2\pi x/L))^\lambda}.
\end{multline}
Integrating this expression over $x$, we obtain
\begin{multline}
\int\limits_0^L dx \langle\cos(\chi(\tau,x)-\chi(0,0))\rangle_0
= L\left(\frac{2\pi^2\xi_c^2}{L^2}\right)^\lambda\\\times\sum\limits_{n=0}^\infty\frac{\Gamma(n+1/2)\Gamma(\lambda+n)}{\sqrt{\pi}\Gamma(\lambda)\Gamma^2(n+1)}\frac{\cosh(4\pi \lambda v\phi_x\tau/L)}{\cosh^{2n+2\lambda}(\pi v\tau/L)},
\label{cosint}
\end{multline}
where $\Gamma(x)$ is the Gamma function.

In order to complete our preparation for the subsequent analytic continuation let us perform the Fourier transformation in Eq. (\ref{cosint}). It is accomplished with the aid of the relation
\begin{widetext}
\begin{multline}
 \int\limits_{-\infty}^\infty d\tau e^{i\omega\tau}\frac{\cosh(4\pi \lambda v\phi_x\tau/L)}{\cosh^{2n+2\lambda}(\pi v\tau/L)}=\frac{2^{2n+2\lambda}L}{4\pi v}\sum\limits_{m=0}^\infty\frac{\Gamma(2\lambda+2n+m)(-1)^m}{\Gamma(2\lambda+2n)\Gamma(m+1)}\left(\frac{1}{\lambda(1-2\phi_x)+n+m-\frac{iL\omega}{2\pi v}}\right.
\\\left.+\frac{1}{\lambda(1+2\phi_x)+n+m-\frac{iL\omega}{2\pi v}}+\frac{1}{\lambda(1-2\phi_x)+n+m+\frac{iL\omega}{2\pi v}}+\frac{1}{\lambda(1+2\phi_x)+n+m+\frac{iL\omega}{2\pi v}}\right)
\end{multline}
\end{widetext}
Substituting this relation into Eq. (\ref{cosint}) and combining it with Eq. (\ref{ccni}) after a simple algebra we arrive at the expression (\ref{piom}).

\end{document}